# Enhanced Autocompensating Quantum Cryptography System

Donald S. Bethune,* Martha Navarro and William P. Risk

IBM Almaden Research Center

650 Harry Road, San Jose, CA 95120-6099

We have improved the hardware and software of our autocompensating system for quantum key distribution by replacing bulk optical components at the end stations with fiber-optic equivalents and implementing software that synchronizes end-station activities, communicates basis choices, corrects errors and performs privacy amplification over a local area network. The all fiber-optic arrangement provides stable, efficient and high-contrast routing of the photons. The low bit error rate leads to high error correction efficiency and minimizes data sacrifice during privacy amplification. Characterization measurements made on a number of commercial avalanche photodiodes are presented that highlight the need for improved devices tailored specifically for quantum information applications. A scheme for frequency shifting the photons returning from Alice's station to allow them to be distinguished from backscattered noise photons is also described.

## 1. Introduction

Quantum cryptography or quantum key distribution (QKD)[1] enables two users, traditionally known as Alice and Bob, to create shared, secret cryptographic keys by transmitting single photons in varying quantum states between their stations over an insecure public channel, for example free-space[2,3,4] or an optical fiber.[5,6,7,8,9,10,11,12,13] Autocompensating QKD systems[8-13] have the feature that they passively and automatically compensate for slow changes in the optical properties of a fiber link, making them convenient and practical. These systems make use of the fact, first noted by Martinelli,[14] that light propagated through an optical fiber and reflected back using a Faraday mirror, returns orthogonally polarized with an accumulated round-trip phase that is independent of the initial polarization state. Gisin and co-workers at the University of Geneva[8-10] first described a system based on this principle and demonstrated it using a 3-Faraday mirror system. A 2x2 fiber coupler and two Faraday mirrors were used to split arbitrarily polarized pulses of light into twin replicas that Bob sent to Alice. She in turn reflected them with a third Faraday mirror, attenuated them to the single-photon level, and imparted a differential phase between the pulses to carry quantum information back to Bob. Bob then read out the phase information interferometrically by recombining the pulses in the original 2x2 coupler and observing from which branch of the coupler the photon emerged.

Intrigued by Martinelli's Faraday orthoconjugation work, we pursued an alternative autocompensating quantum cryptography system that used a polarizing beamsplitter to split a pulse into equal orthogonally polarized components.[11] One component was delayed and then both were sent to Alice. She attenuated and differentially phase-shifted the two components by an amount $\Delta\phi_A$ using a fast phase modulator, and Faraday-reflected them back to Bob. They returned, still orthogonally polarized, but rotated 90°. Bob could then alter the differential phase shift between the orthogonal components using his own fast phase modulator to shift the first component by $\Delta\phi_B$ before it recombined with the second on the original polarizing beamsplitter. The net differential phase $\Delta\phi = (\Delta\phi_A - \Delta\phi_B)$ controlled the polarization state of the recombined beam, allowing the quantum information to be read out using polarization analyzing bulk optics.

Ribordy et al.[12] reported an elegant upgrade of the Geneva group's system, which also used a polarizing beamsplitter to send orthogonally polarized, time-separated pulses to Alice. In their case, rather than combining the returning single-photon pulses orthogonally and decoding the resulting polarization state, polarization controllers were used to transform the pulses to a common (arbitrary) polarization state, so that they could be interferometrically combined using a normal 2x2 fiber coupler to give phase-shift sensitive routing of the photons. An adaptation of this system for 1.55 µm operation has also been reported.[13]

In this paper we report the performance of an upgraded version of our autocompensating quantum cryptography system. In addition, data characterizing the performance of a variety of avalanche photodiodes as single-photon detectors for telecom wavelengths are presented. Finally we describe initial tests of a scheme for avoiding backscattering noise based on the idea of frequency shifting the photons returning from Alice's end station. Because of this frequency labeling, narrow-band filters at the detectors can be used to discriminate against backscattered photons.

## 2. All-fiber System

Our modified system is shown in Figure 1. Generally, the custom bulk-optic components used in our original implementation[11] have been replaced with polarization-maintaining (PM) fiber-optic components. At Bob's station, a laser pulse is split into two equal-amplitude parallel-polarized pulses using a 2x2 fiber coupler, similar to the Geneva group's later arrangement,[12] but here using a variable, polarization-preserving coupler. The use of all polarization maintaining components in Bob's delay loop enhances the stability of the system and eliminates the need for polarization adjusters in the phase-sensitive parts of the optical path. The one polarization adjuster in the system precedes the variable coupler, and is set to maximize transmission through the guided-wave phase modulator in Bob's delay loop ($Mod_B$), which was fabricated by the annealed proton-exchange (APE) process and hence transmits only one linear polarization.  The phase modulator output is connected to one input of a 2x2 fiber-optic polarizing beamsplitter[15] (PBS1, used here as a 2x1 coupler). The other variable coupler output fiber is connected to the second PM input fiber of PBS1 using a modified FC/PC connector that aligns the fast axis of one fiber with the slow axis of the other.  PBS1 thus combines the split pulses so that they leave Bob's station orthogonally polarized and separated in time by ~30 ns.

The main improvement at Alice's station is the replacement of the bulk-optic polarizer with a 2x2 fiber-optic polarizing beamsplitter similar to Bob's (PBS2).  Alice's polarizing coupler allows her to impart equal phase shifts to *both* polarization components of an *arbitrarily polarized* arriving pulse using a single-polarization phase modulator ($Mod_A$) similar to Bob's.  After PBS2 separates them, the two components travel around the PM fiber loop in opposite directions and meet in the modulator with the same polarization (since one component enters the loop after Faraday reflection and the other before). They are given a common phase shift by the modulator, and then retrace each other's paths and recombine at PBS2 into a single pulse that heads back towards Bob.  The overall round-trip loss at Alice's station is ~32 dB, and the subsequent one-way transmission loss from Alice's station to Bob's detectors, including fixed losses and fiber attenuation, is ~11 dB.  Alice encodes information in the pulse pairs by applying a phase shift $\phi_A = -\pi/2, 0, \pi/2$, or $\pi$ to the second pulse of each pair using her modulator ($Mod_A$), corresponding to (basis,bit) pairs (1,0), (0,1), (1,1), and (0,0), respectively.  Bob selects his decoding basis (0 or 1) by applying a phase $\phi_B=0$ or $-\pi/2$ to the first pulse of the returning pair as it travels through the delay loop.  The pulses arrive back at the initial beamsplitter identically polarized and with a total differential phase shift $(\pi+\Delta\phi)$, where $\Delta\phi = (\phi_A-\phi_B)$ and the additional $\pi$ phase shift is due to the 90° Faraday rotation.  For $\Delta\phi=0$ or $\pi$ photons are deterministically routed to $D_0$ or $D_1$, respectively, while for $\Delta\phi$ an odd multiple of $\pi/2$ photons arrive randomly at either detector with equal probability.  The variable coupler was initially set to precisely balance the amplitudes of the recombining pulses by nulling the count rate of $D_0$ with $\Delta\phi=\pi$.  Subsequently the split ratio was quite stable and needed only infrequent slight adjustments.

The detection electronics are the same as described in our previous report (Ref. 11, and further details below).  For each channel a Fujitsu FPD5W1KS InGaAs avalanche photodiode (APD) at T=118 K is reverse-biased to ~30 V DC.  Pulses of 1.5 ns duration and 3.3 V amplitude are superimposed on the DC voltage in order to periodically raise the total bias just above the reverse breakdown voltage.  The DC bias was adjusted to give detector quantum efficiencies $\eta_B$~20 %, so with $\mu$=0.3 photon/pulse at 1 MHz, the total detection rate was ~5x10$^3$ /s.  Fig. 2 shows the count rates for $D_0$ and $D_1$ versus the time delay between the detector bias pulses and the arrival of the optical signal for modulator settings $\Delta\phi = 0$ and $\pi$. The detection window has a FWHM of ~750 ps.  The switching contrast ratios achieved for the two phase settings averaged about ~650:1, corresponding to a BER contribution of only 0.15 %. This routing contrast is nearly 20 times better than that achieved with our previous optical mini-bench, which used bulk polarization analyzing optics. The off-peak count rates are due to dark counts (~40 /s for each detector) and backscattered 1.31 μm photons (~15 and 25 Hz for $D_0$ and $D_1$, respectively, for 0.3 photon/pulse returning pulses). The total BER for our 10 km link at μ=0.3 is thus ~1.5 %.

## 3. Key generation results and security considerations

The results for key generation with the full BB84 protocol over the 10 km link are summarized in Figures 3a, which shows the raw, error-corrected, and privacy-amplified bit rates and corresponding BER's for varying μ. Three stages of block-parity comparison error correction (with block row lengths 40/50/60) reduced the BER to ~0.2% or less, and then random-subset parity comparisons were used to eliminate the sparse remaining errors. All errors could be eliminated at a cost of 10-15% of the raw bits for the 10 km link. Privacy amplification[1,2] was implemented by having Alice and Bob compute the parities of M additional randomly chosen sets of N/2 of their N shared, error-corrected bits. M was chosen to be [N-(L+S)], where L is the estimated maximum number of the N bits known by Eve, and we chose S=30 to dilute Eve's information about the final key (proportional to $2^{-S}$) to a negligible level.

For benchmarking purposes we used the simple estimate for the fraction of the raw bits potentially known to Eve given in BB84[1]: (2B+μ+5v), where B is the BER, μ is the average number of photons/pulse, and v is a variance for the first two quantities. This estimate makes the assumption that Eve can delay measuring beam-split pulses until the bases are revealed, and approximates the maximum probability that Eve will detect a photon for a given pulse, [1-exp(-μ)], as μ. At the maximum of the rate curve in Fig. 3a, about 0.6 privacy amplified bits per raw key bit are obtained, corresponding to a maximum rate of privacy amplified key generation of ~1.5 kb/s.

The leak estimation formula in our software can readily be changed to use any appropriate upper bound for information leakage. The value μ=0.3 photons/pulse used in taking the data in Fig. 3 is close to optimum under our conditions *if* only simple beamsplitter and read-and-replace eavesdropping attacks are considered. This moderately high value of μ is then favorable because it gives a higher raw data rate and lower BER, leading to higher error correction efficiency. The maximum estimated information leak rate to Eve is then dominated by the beamsplitting attack. Recent studies of the security of QKD [16, 17, 18, 19, 20, 21, 22, 23], however, have shown that granting Eve measurement abilities well beyond current technology, but consistent with known laws of physics, makes her a much more formidable opponent. In References 21 and 22 for example, it is assumed that Eve can measure the photon number of every pulse, forward all multiphoton pulses to Bob undisturbed over a lossless channel keeping exactly one photon from each for herself, and store and analyze these photons after the bases are revealed. For the multiphoton pulses, she switches Bob's detector efficiencies to unity. Finally, she intercepts, reads, and tries to repeat a certain fraction of the single-photon pulses. For these pulses Eve induces a 25% error rate, but any remaining single-photon pulses and the error-free multiphoton pulses could dilute this rate to a level that would be acceptable to Alice and Bob. Such a powerful attack puts a very severe restriction on the allowable pulse size, since to obtain any secure key data under such conditions the ratio of multiphoton to single photon pulses must be sufficiently small that a key based only on single-photon pulses received *in excess of the multiphoton pulses transmitted* can be obtained. In particular, for Poisson distributed pulses and neglecting dark counts and error correction, Lütkenhaus[21] shows that under the above assumptions the optimum rate of secure key generation is obtained if $\mu \sim \eta_B \eta_T$, where $\eta_B$ is Bob's detection efficiency and $\eta_T$ the link transmission probability. For this choice of $\mu$ the gain G, defined as the number of secure bits obtained per transmitted pulse, is $G \sim \frac{1}{4} (\eta_B \eta_T)^2$. For our 10 km link this bound is ~60 bits/sec (at $\mu$=0.016). When dark counts and error correction are taken into consideration the net rate is reduced to near zero.

A recent analysis by Gilbert and Hamrick[23] argues that the analysis of Ref. 21 is somewhat overly conservative, particularly in its handling of 2-photon pulses. Rather than granting Eve a bit for each *transmitted* 2-photon pulse, Ref. 23 points out that Eve cannot obtain the exact photon state by analyzing 2-photon pulses (and thus cannot use a read/repeat attack without inducing errors), so her best attack is to split the 2-photon pulses, keeping one of the photons and analyzing it after the bases are announced. Crucially, only those pulses for which Bob detects the second photon contribute to Eve's knowledge of the key, and because the no-cloning theorem prohibits faithful amplification of those photons, Ref. 23 argues that the number of 2-photon pulses Bob detects is necessarily proportional to his detector quantum efficiency, $\eta_B$. Following this analysis, fewer of the 2-photon pulses give Eve useful information, and therefore fewer bits must be sacrificed to obtain adequate privacy amplification. With the prescription of Ref. 23 for privacy amplification, the peak secure key generation rate for our 10 km link would be ~200 bits/sec at $\mu \sim 0.1$.

We also tried a 20 km SMF28 link, with results summarized in Figure 3b. For the longer fiber, increased attenuation (~3 dB, one way) and slightly increased timing jitter reduced the raw bit rate with $\mu$=0.3 /pulse to ~$2 \times 10^3$/s. Under these conditions, the BER contribution due to dark counts increased to ~2%, and that due to increased 1.3 $\mu$m backscattering rose to ~2.6%, giving a total BER of ~4.6%. At this BER level, error correction (using block row lengths 12/18/27) consumed 35-40% of the bits. The lower raw bit rate and reduced error correction and privacy amplification efficiencies produced a BB84 privacy amplified key generation rate of ~150 bits/sec. With the privacy amplification prescription of Ref. 23 this rate drops to a few bits/sec, while with that of Refs. 21 and 22 no net bits are obtained. At longer distances it will clearly be necessary to increase the system efficiency by some combination of reducing losses, backscatter and dark counts and improving the detector quantum efficiency.

## 4. Single-photon detector characterization

The need to correct all key errors and implement privacy amplification in quantum cryptography systems makes it critically important to have high efficiency, low BER detectors. A number of authors have studied the properties of APDs based on low-bandgap materials such as Ge and InGaAs for single photon detection.[24, 25, 26, 27, 28, 29, 30, 31, 32] In general low temperature operation is required to reduce dark count rates to acceptably low levels - typically 77 K for Ge devices and from 100-238 K for InGaAs devices. In hopes of finding detectors superior (or at least comparable!) to the Fujitsu devices described above, we characterized InGaAs and Ge APDs from several manufacturers as single-photon detectors for 1.3 and 1.55 µm light (see Table I). [33 34 35] All APDs were mounted on a temperature regulated copper plate. Quantum efficiencies and dark count rates were measured for temperatures ranging from 90 K to above 200 K. A schematic of the detector characterization setup is shown in Figure 4 [36 37 38 39 40]. As in our quantum cryptography system described above, the diodes were DC biased below their reverse breakdown voltage, $V_{br}$, and pulse biased above $V_{br}$ for 1.5 ns intervals at a rate of 1 MHz. Capacitive transients coupled through the APD were canceled using equal lengths of 50 Ω cable attached via subminiature SMA tees to the anode and cathode of the APD. This arrangement has the advantage of reducing the average current through the device, which in turn lowers the amount of trapped charge and probability of after-pulsing (see Ref. 10 for further details). Labview[41] routines were developed to control the bias voltage, sample temperature, and counting electronics.

The performance of the APDs is summarized in Fig. 5, which shows the dark counts per bias pulse and quantum efficiencies obtained as the DC bias voltage applied to the device was varied. All of the APDs were InGaAs devices with the exception of #6, which was Ge. The highest performing devices were Fujitsu type FPD5W1KSF. Two of these (#'s 7 & 8) are the devices used in the cryptography experiments, with the typical operating range indicated by the dashed circle. We note that nominally identical devices of later vintage (#'s 2 & 5) exhibit drastically different behavior, showing two orders of magnitude higher dark rates for QE's ~15%. This is perhaps not surprising, since even though all of the devices have excellent room temperature characteristics, none was designed for low temperature operation and their low temperature characteristics are neither controlled for nor documented by the manufacturers. We also note that even though the NEC Ge device has a relatively high dark rate at low bias, that rate remains approximately constant as the bias increases while the QE rises steeply over a few tenths of a volt. Thus for QE's>15%, the Ge device is second only to the best Fujitsu diodes in performance.

In order to compare different APDs taking into account the trade off between obtaining a high detection rate at high bias voltage and a low dark-count rate at lower bias, we devised an ad hoc figure of merit, **K**=(S/D)xQE =$10^3 \cdot QE^2/D$ (where S/D is the signal-to-dark count ratio with µ=0.1 and a 1 MHz pulse rate, and QE is the quantum efficiency in %).[42] The variation of **K** with DC bias voltage at various temperatures for one of the best Fujitsu APDs (#9) is shown in Figure 6. Data for both 1.3 and 1.55 µm are shown for T in the range 108–190 K. At 1.3 µm, the exponential fall of dark counts with temperature leads to values of **K** exceeding 24,000 at 118 K (S/D~1200, QE~20%). Below 118 K the dark counts begin to increase again, presumably because of increasing lifetimes for electrons in shallow traps. The values of **K** for 1.55 µm are lower, particularly below 160 K, since **K** is proportional to $QE^2$ for fixed dark counts, and the 1.55 µm absorptivity falls with temperature reducing the QE of the device. A maximum **K** value of ~2400 (S/N=270, QE~9%) was attained at 160 K for 1.55 µm light.

Figure 7 summarizes the temperature dependence of the maximum **K** values for a number of devices. Despite their similar performance at room temperature, the **K** factors for the devices are spread over 3 orders of magnitude, and here too it is seen that even devices of the same type from the same manufacturer vary widely in their characteristics.

As have other authors on this subject, we conclude that there is no way to select devices for single-photon counting based on their room temperature performance. Generally there is a strong need for APDs specifically tailored for single-photon detection at telecom wavelengths for applications to quantum cryptography. This need is particularly urgent in view of the fact that as telecom technology advances, the requirements of the industry are rapidly changing and products for standard applications may become unsuitable for low temperature operation (as seems to be the case with later vintage Fujitsu APDs) or worse, may become unavailable altogether (as is the case for the EG&G diodes, the NEC NDL5151P1 Ge APD, and the Fujitsu FPD5W1KSF, which has now been integrated with a GaAs preamp and is no longer offered as a discrete unit). On the positive side, the fact that some devices not specifically designed for photon counting perform reasonably well (for example, the EPM 239 AA SS [43] according to two recent reports[31, 32]) gives us confidence that with deliberate engineering effort superior devices with both higher efficiency and lower noise can be made.

## 5. Backscatter suppression by frequency shifting returning photons

Our autocompensating system runs with a continuous stream of outgoing and returning pulses. In such a round-trip system, Raleigh backscattering of photons from the outbound light eventually becomes intolerable as the fiber length increases. In our system it already limits system performance with a 20 km link. An intriguing possibility for reducing noise due to backscattering is to distinguish the signal bearing photons from scattered light by shifting the frequency of the photons returning from Alice. With high efficiency, narrow band filters, Bob could then block the scattered noise photons and detect the information bearing photons with a greatly reduced background. We have carried out an initial investigation of this approach using the setup shown schematically in Figure 8. The conflicting desires for high transmission through a narrowband filter and short pulse lengths to minimize detector on-time and dark counts dictates the use of near Fourier-transform limited pulses. These were generated at Bob's station using a narrowband cw-laser/amplitude modulator combination to produce 1 ns pulses with a bandwidth of ~1 GHz. At Alice's station the photons were sent through a 1 GHz acousto-optic Bragg cell. In double-passing the cell, photons are up-shifted in frequency by 2 GHz. Bob uses a custom Innovative Fibers[44] fiber-optic filter consisting of a Bragg-grating phase-shifted at the center by half the grating period. This design produces a narrow transmission band (FWHM ~1 GHz) within a broader reflection band (FWHM ~175 GHz)[45, 46, 47] and can thus be used to transmit frequency-shifted pulses and block unshifted, scattered light. The on-peak transmission of the filter is >90% while the transmission 2 GHz off-peak is ~7%.

Measurements of the routing contrast similar to those depicted in Figure 2 were made over the 20 km fiber link to assess whether frequency shifting the returning light disrupts the Faraday autocompensation. The contrast was found to be ~99% with the 2 GHz shifter operating, indistinguishable from the value obtained using the straight-thru port of the Bragg cell with the shifter off. This favorable result reflects the fact that the round-trip differential phase shift $\Delta\phi_S$ introduced by frequency shifting the returning light by $\Delta f$ is expected to be of order $(2\pi\Delta f \cdot PMD \cdot L^{1/2})$, where PMD is the polarization mode dispersion coefficient for the fiber and L is the fiber length. Corning SMF-28 is specified to have PMD $\leq 0.2$ ps/km$^{1/2}$ for a single fiber, leading to an estimate of $\Delta\phi_S$ ~13 mrad for a 20 km length. Since the fraction of misrouted photons is $\sim\Delta\phi_S^2$, this gives a negligible loss of fringe contrast. This estimate also suggests that frequency shifts up to ~20 GHz would degrade the switching contrast by less than ~2%, indicating that the autocompensation is relatively robust against frequency shifts for low PMD fiber.

Figure 9 shows the detection rate vs. tunable laser wavelength measured with this setup. The data can be fit as the sum of three Lorentzians: peak 1 is the desired peak for returning shifted pulses and peaks 2 and 3 are unshifted peaks due to leakage of narrowband CW light through the amplitude modulator and backscattering of outgoing pulses. In addition, a broadband background with intensity ~1/3 of the shifted peak intensity due to backscattering of laser super-fluorescence underlies the spectrum. To make this arrangement work for quantum cryptography, this broad background would need to be eliminated using a band-pass filter in the laser output line. The on-off ratio of the amplitude modulator (~27 dB) would ideally also be improved, but since the backscattered modulator leakage is narrow and unshifted, the filter at the detector mostly eliminates it. We conclude that by shifting the signal photon frequency by 2 GHz, the backscatter/signal ratio can be improved by roughly an order of magnitude with this scheme, but at the cost of requiring a more highly controlled light source, matched filters at the source and detector, and somewhat reduced transmission to the detectors. With larger frequency shifts it would be possible to obtain even better discrimination.

An alternative method for avoiding backscattered light noise has been demonstrated by Ribordy et al,[12] who used intermittent transmission of pulses and a delay line behind Alice's attenuator to allow detection of the signal photons during intervals when essentially no background light arrived at the detectors. The drawback to this approach is that the system duty factor is reduced by a factor $[L_D/(L_T+L_D)]$, where $L_T$ is the link length and $L_D$ is the delay line length. However, since Alice must attenuate the pulses before sending them back to Bob anyway, reasonably long delay lines are feasible, so the loss of key generation rate due to the reduced duty factor can be kept to a reasonable level.

## 6. Conclusions

Replacement of bulk optic components with fiber optic counterparts in our autocompensating quantum cryptography system reduced the bit error rate over a 10-km link from ~5% to ~1.5%. This reduced error rate allows error correction and privacy amplification to be carried out with significantly improved efficiency, with the result that the generation rate of final key was increased from about ~200 bits/s in our earlier system to ~1.5 kbits/s in this all-fiber version, using BB84 privacy amplification. The improved performance also enabled us to perform quantum key distribution over a 20-km link at a reduced rate due to increased backscattering of outgoing light and higher attenuation. The use of polarization-preserving fiber components in the upgraded system, particularly in the delay loop at Bob's station, combined with the autocompensating design, results in a very stable system requiring little adjustment during operation.

Our work characterizing various APDs as single-photon detectors shows the clear need for the development of better quantum detectors tailored specifically to quantum cryptography and other quantum information applications. We particularly note that some of the most suitable devices we found, imperfect as they may be, are no longer even available.

Finally, the scheme we described for reducing the noise due to backscattered photons by frequency shifting the returning photons was shown to give an order of magnitude reduction in the backscatter noise without degrading the interferometric contrast. This would be sufficient to allow autocompensating systems with continuous pulse trains to be used with 20 km or longer fiber links.

*D.S. Bethune (bethune@almaden.ibm.com) and W.P. Risk are with the IBM Almaden Research Center, 650 Harry Road, San Jose, California 95120-6099. M. Navarro, now with the Department of Physics, University of New Mexico, Albuquerque, NM 87131, was at IBM at the time of this research.

**Avalanche Photodiodes Tested**

| # | Manufacturer | Model | Serial | Material | $V_B(V)$ | $K_{max}$ | $T_{max}(K)$ |
|---|---|---|---|---|---|---|---|
| 1 | Fujitsu[33] | FPD5W1KSF | DY-5 | InGaAs | NA | 472 | 108 |
| 2 | Fujitsu | FPD5W1KSF | 544 (A) | InGaAs | 53.8 | 24520 | 118 |
| 3 | Fujitsu | FPD5W1KSF | 553 (B) | InGaAs | 52.8 | 22000 | 118 |
| 4 | Fujitsu | FPD5W1KSF | 501 (C) | InGaAs | 54.7 | 22000 | 118 |
| 5 | Fujitsu | FPD5W1KSF | EY-44 | InGaAs | 47.4 | 210 | 108 |
| 6 | NEC[34] | NDL5551P1 | 287 | InGaAs | 66.2 | 1038 | 130 |
| 7 | NEC | NDL5551P1 | 291 | InGaAs | 64.4 | 556 | 130 |
| 8 | NEC | NDL5151P1 | 270 | Ge | 34.3 | 2192 | 77 |
| 9 | EG&G[35] | C30644 EJT-07 | 1661 | InGaAs | 44 | 43 | 240 |

Table I

Subset of tested APDs corresponding to data shown in Figure 5. All devices were used with the pulse-bias setup shown in schematic in Figure 4. $K_{max}$ is the global maximum value obtained for the product of the signal-to-dark ratio (at 0.1 photon/pulse and 1 MHz pulse rate) and the quantum efficiency (in %) for all measured DC biases and temperatures. $T_{max}$ is the temperature at which $K_{max}$ was attained.

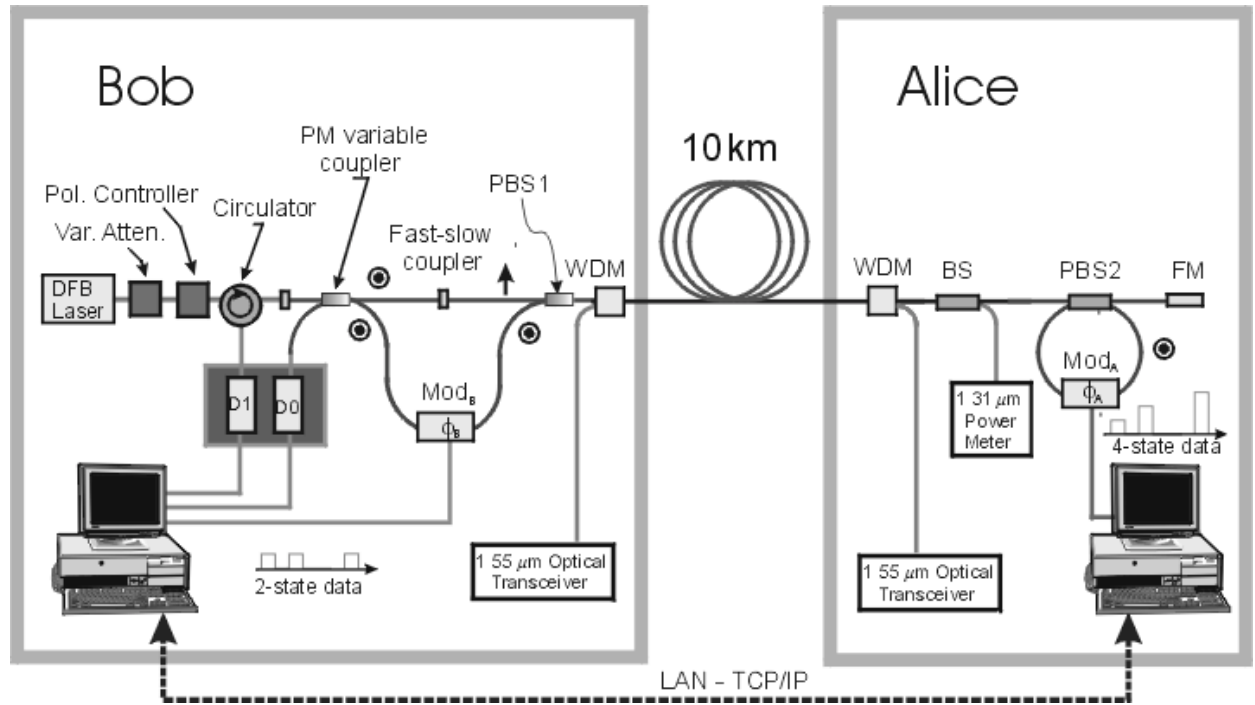

Figure 1. DFB laser emits 50 ps pulses at 1 MHz rate. PBS1,2 – fiber-optic polarizing beamsplitters; WDM – wavelength division multiplexers for 1.31 and 1.55 μm; FM – Faraday Mirror; BS – fiber-optic beamsplitter; $Mod_{A,B}$ – APE LiNbO$_3$ phase modulators; $D_{0,1}$ – Fujitsu FPD5W1KSF InGaAs avalanche photodiodes at 118 K. The fast-slow coupler is an ordinary FCPC coupler with the two key slots rotated 90º apart.

**Figure 1**

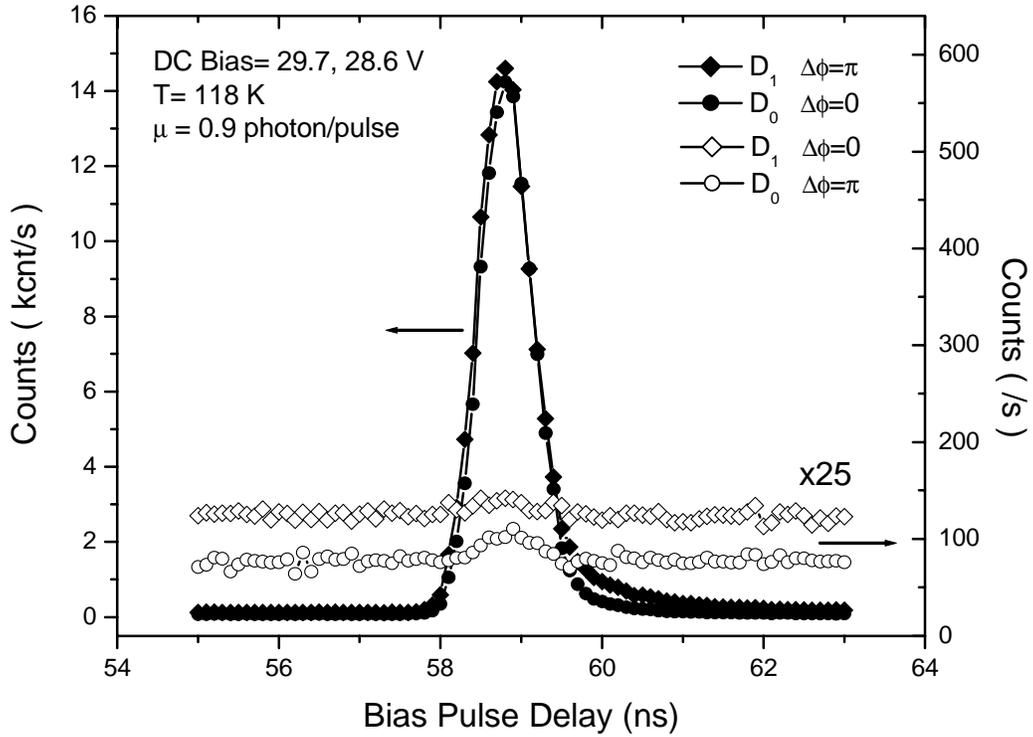

Figure 2. Count rates vs. detector bias pulse delay for relative phase shifts $\Delta\phi = 0, \pi$. Contrast ratios are 29.7 and 26.9 dB, respectively. The detector dark rates (at 1 MHz bias rate) are ~40 /s. Backscattered 1.31 μm light also contributes to the off-peak counting rates. The pulse intensity was increased to 0.9 photons/pulse for these measurements to increase the count rate for the switched-off channel (normally small compared to the dark count rates).

**Figure 2**

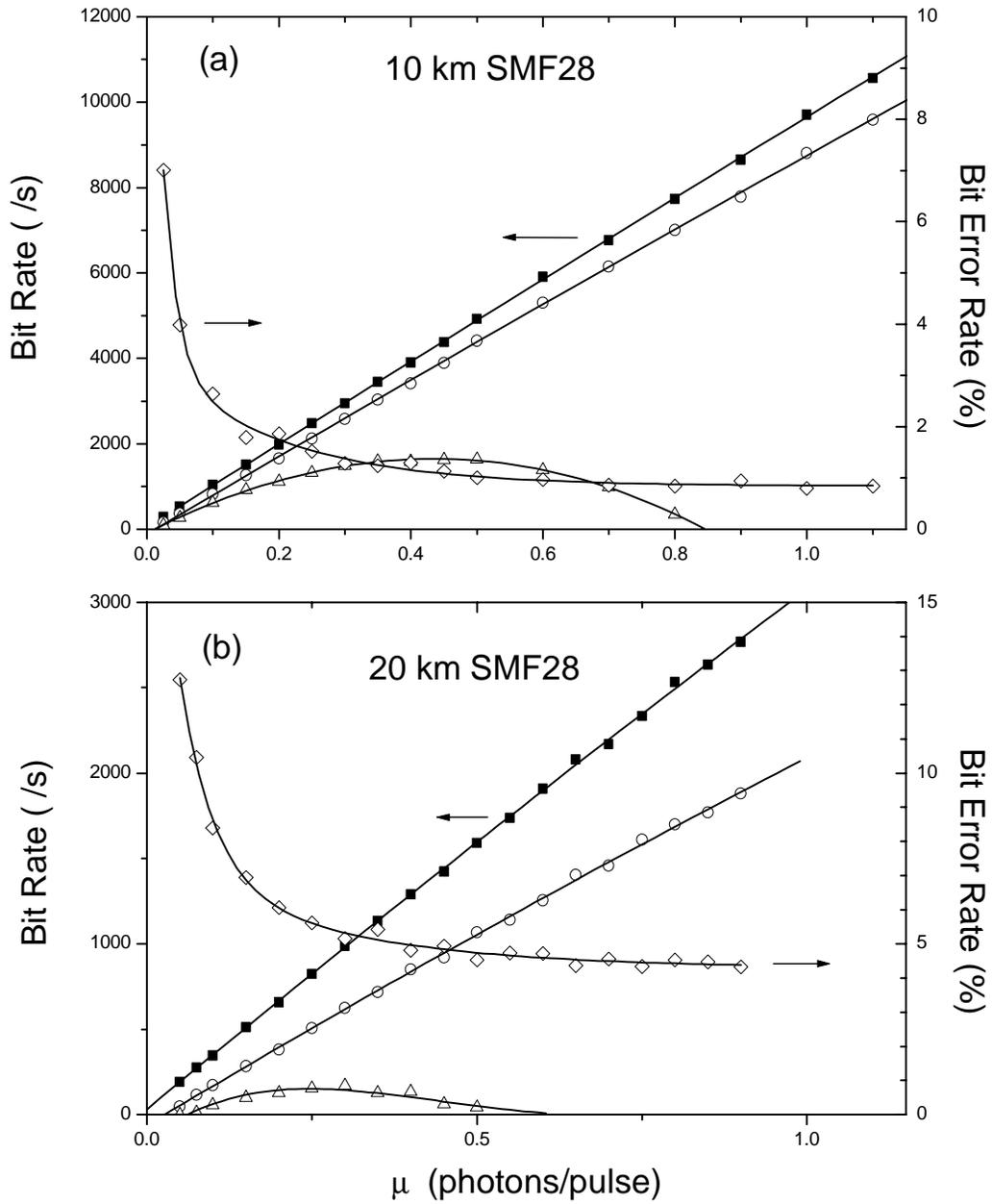

Figure 3. a) Data for 10 km fiber; b) data for 20 km fiber. Raw (■), error corrected (O) and privacy amplified (Δ) bit rates, and measured bit error rates (◊), vs. the mean number of photons/pulse, $\mu$, leaving Alice's station. The leakage rate to Eve is estimated using the original BB84 formula (see text). All curves are simple fits provided as guides to the eye.

**Figure 3**

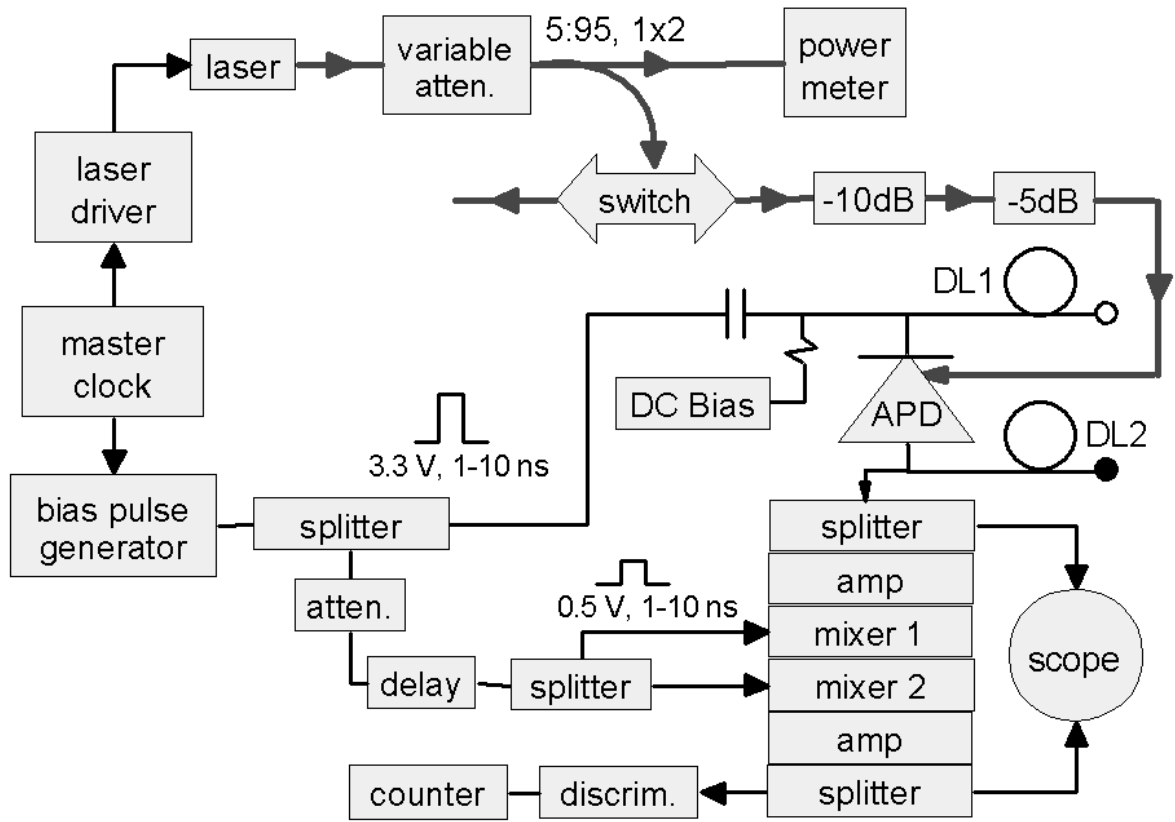

Figure 4. Single-photon counting APD test set-up. Master clock is a SRS-DG535 [36] that triggers an Avtech[37] AVO-9C laser driver and the HP8131A bias pulse generator[38]. Minicircuits[39] ZFRSC42 or ZFRSC2-2 splitters, ZFL2000 amplifiers, and ZEM2B mixers are used for the APD output electronics, feeding a Lecroy[40] 4608c discriminator. DL1 and DL2 are matched 6.5 ns delay lines attached by SMA tees to the anode and cathode of the APD. DL1 is unterminated, while DL2 is shorted.

**Figure 4**

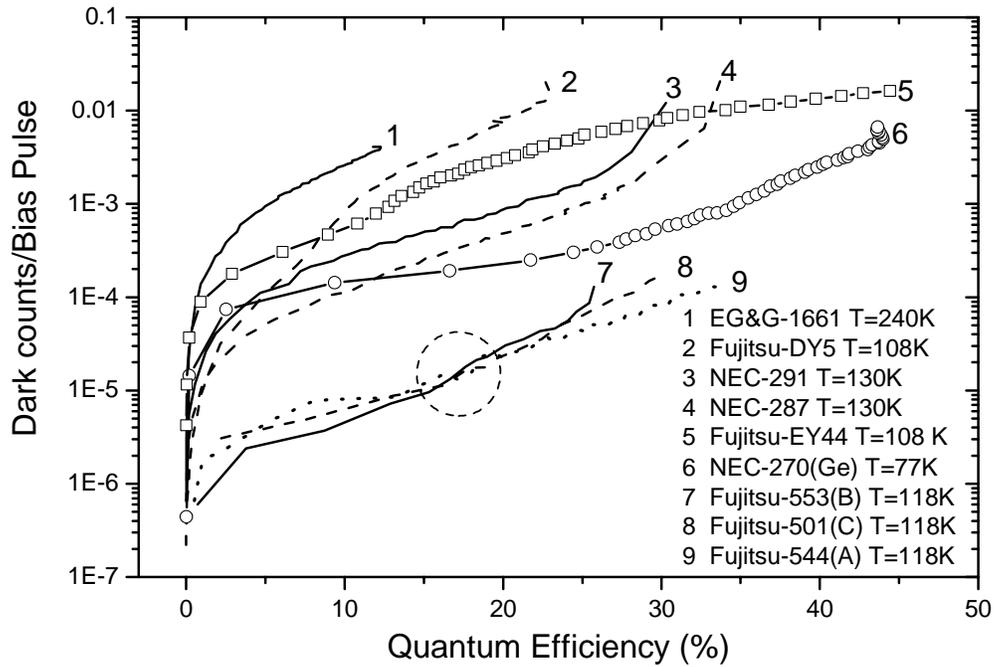

Figure 5. Dark counts/bias pulse vs APD QE for various APDs with 1.31 μm light at their optimum temperatures. The DC bias voltage was varied over ~3 V in 50 mV steps to trace the curves. APDs 7 and 8 are the devices used in the quantum cryptography experiments, with typical operating points within the dashed circle.

**Figure 5**

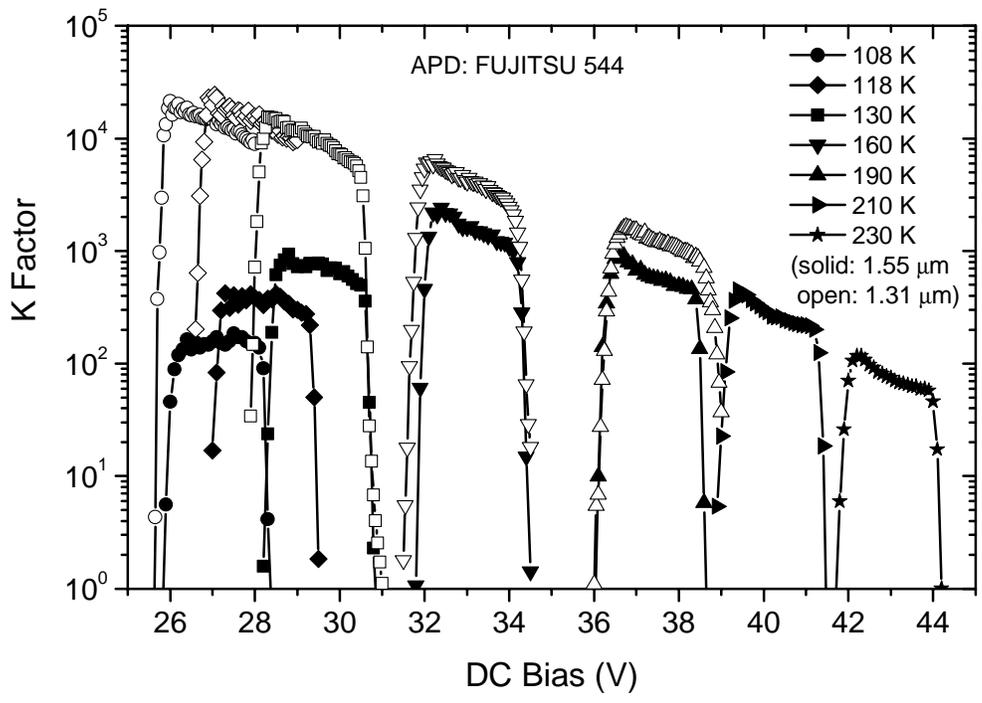

Figure 6. **K** values for Fujitsu FPD5W1KSF APD (#2 in Table) vs. DC bias voltage for various temperatures (see text).

**Figure 6**

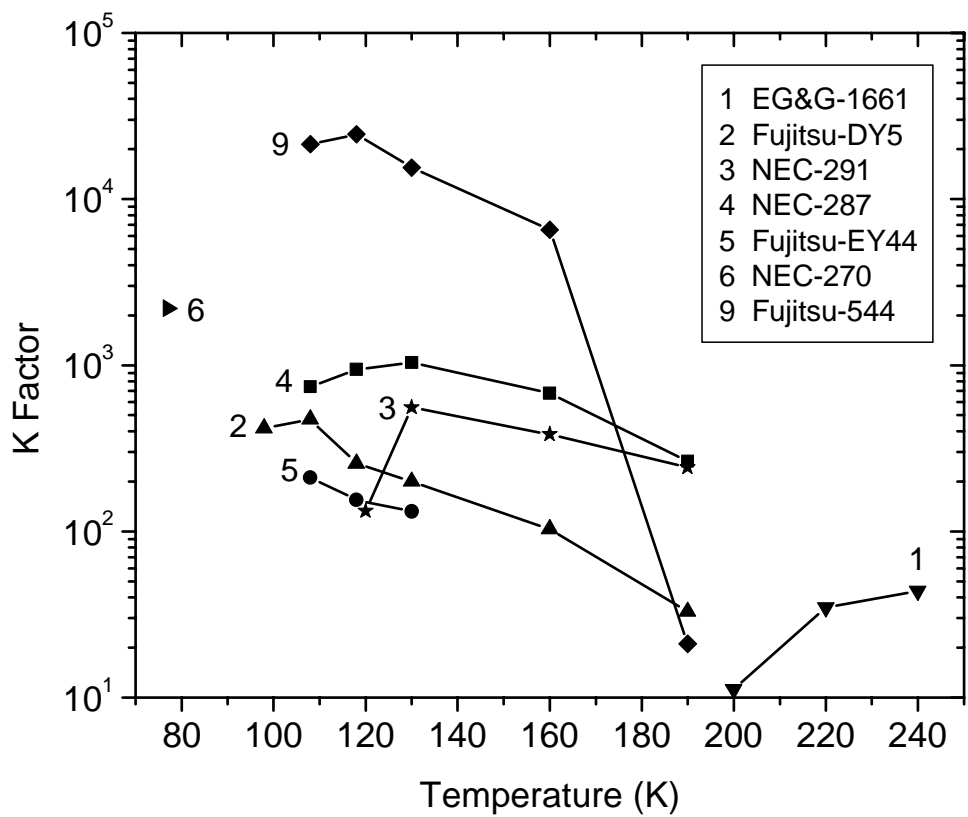

Figure 7. Maximum **K** values attained vs. T for various detectors. The NEC NDL5151P1 Ge APD (#8 in Table) was only tested at 77 K.

**Figure 7**

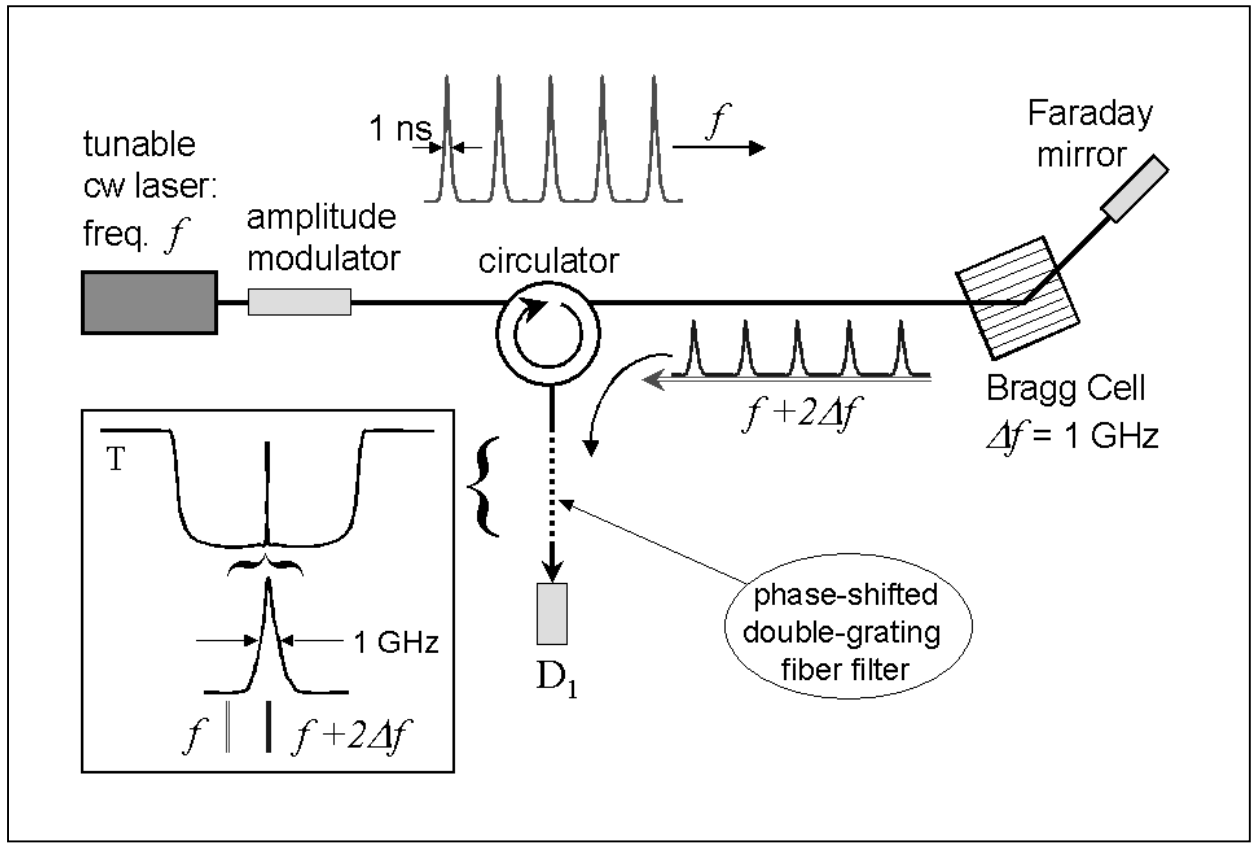

Figure 8. Schematic of setup for frequency shifting photons at Alice's station using a 1 GHz acousto-optic Bragg cell. In double-passing the cell, the photons are shifted up in frequency by 2 GHz. The narrowband cw laser/amplitude modulator combination produces 1 ns pulses with near-Fourier-transform-limited bandwidth. Bob uses an Innovative Fibers[44] phase-shifted double Bragg grating fiber-optic filter with FWHM of ~1 GHz to transmit the frequency shifted pulses and reject unshifted, scattered light. The filter blocking band is ~ 175 GHz wide.

**Figure 8**

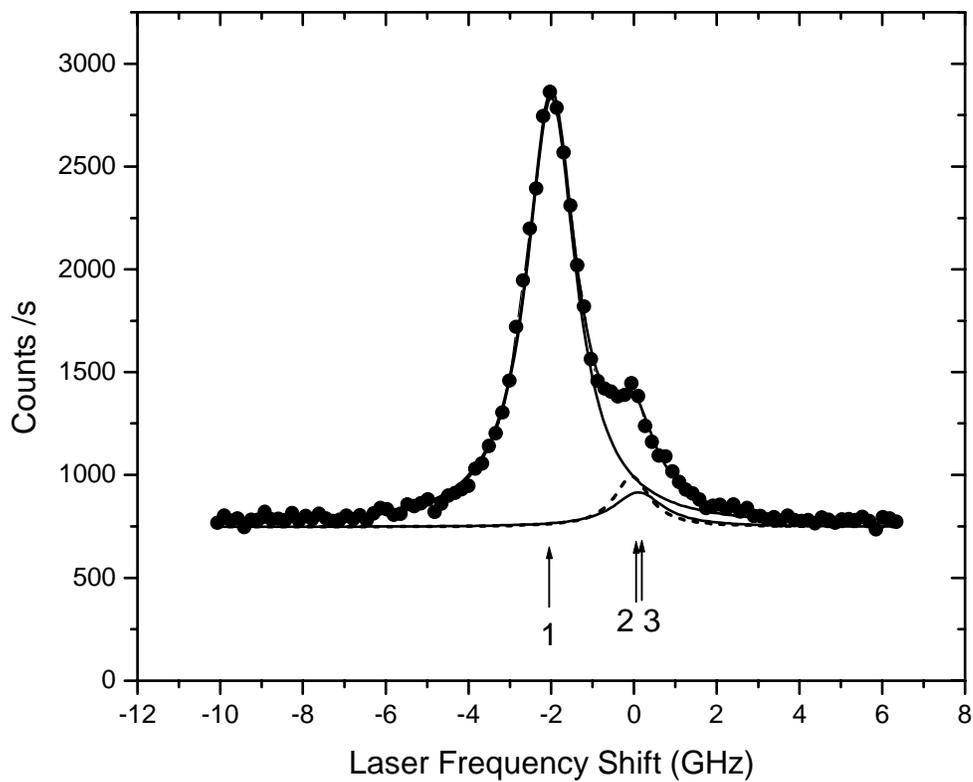

Figure 9. Detection rate vs. tunable laser wavelength with 2 GHz acousto-optic shifter at Alice's station. Data is fit with three Lorentzians: peak 1 is the desired peak for returning shifted pulses, while unshifted peaks 2 (dashed curve) & 3 (solid curve) are due to leakage of narrowband CW light through the amplitude modulator and backscattering of outgoing pulses, respectively. In addition, a broadband background of ~700 counts/sec due to backscattering of laser superfluorescence underlies the spectrum.

**Figure 9**